\documentclass[namedreferences]{solarphysics}

\usepackage[hyperref,optionalrh]{spr-sola-addons} 
\usepackage{graphicx}        
\usepackage{color}           
\usepackage{breakurl}        

\newcommand{\aap}{    {\it Astron. Astrophys.}}

\newcommand{\apjl}{   {\it Astrophys. J. Lett.}}

\chardef\us=`\_

\begin{document}
\begin{article} 
\begin{opening}
\title{Annual Cosmic Ray Spectra from 250~MeV up to 1.6~GeV from 1995\,--\,2014 Measured With the Electron Proton Helium Instrument onboard SOHO}

\author[addressref=aff1,email={kuehl@physik.uni-kiel.de}]{P. K\"uhl}	
\author[addressref=aff2]{R. G\'{o}mez-Herrero}
\author[addressref=aff1]{B. Heber}

\address[id=aff1]{Institute for experimental and applied physics, University Kiel, 24118 Kiel, Germany}
\address[id=aff2]{Space Research Group, University of Alcal\'{a}, E-28871 Alcal\'{a} de Henares, Spain}

\runningauthor{P. K\"uhl et al. 2015}
\runningtitle{Annual Cosmic Ray Spectra Measured with SOHO/EPHIN}

\begin{abstract}
The solar modulation of galactic cosmic rays (GCR) can be studied in detail by examining long-term variations of the GCR energy spectrum (e.g. on the scales of a solar cycle). With almost 20 years of data, the \textit{Electron Proton Helium INstrument} (EPHIN) onboard the \textit{SOlar and Heliospheric Observatory} (SOHO) is well suited for this kind of investigation.
Although the design of the instrument is optimized to measure proton and helium isotope spectra up to 50 MeV nucleon$^{-1}$, the capability exists to determine proton energy spectra from 250~MeV up to above 1.6~GeV. Therefore we developed a sophisticated inversion method to calculate such proton spectra. The method relies on a GEANT4 Monte Carlo simulation of the instrument and a simplified spacecraft model that calculates the energy-response function of EPHIN for electrons, protons and heavier ions. 
For validation purposes, proton spectra based on this method are compared to various balloon missions and space instrumentation. As a result we present annual galactic cosmic ray spectra from 1995 to 2014. 

\end{abstract}


\end{opening}
\section{Introduction}
\citet{hess1912} discovered the evidence of a very penetrating radiation later called cosmic rays, coming from outside the atmosphere. When \citet{par58} described the solar wind, theoretical research of cosmic rays began, stimulated by the beginning of \textit{in-situ} space observations which have led to over four decades of important space missions, including the Voyager, Ulysses, and -- more recently -- the \textit{Payload for Antimatter Matter Exploration and Light-nuclei Astrophysics} (PAMELA) and the \textit{Alpha Magnetic Spectrometer} (AMS) missions. It is well known that at energies below several GeV the cosmic-ray flux is anti-correlated with the 11-year and 22-year solar-activity cycle due to the solar modulation \citep{heber2006a, heber2006b}. Thanks to the Voyager mission, uncertainties about the modulation volume (\textit{i.e.} the heliosphere) as well as the local interstellar spectrum (LIS), became small \citep{sto13}. Thus the remaining main question to pose is: How do charged particles propagate through the three-dimensional heliosphere and how do their transport and propagation vary with the particles' energy and the solar activity? \newline
The transport of cosmic rays in the heliosphere can be described by Parker's transport equation \citep{par65}. Physical processes that determine the measured flux at 1 AU are:
i) Outward convection caused by the radially directed solar-wind velocity.;
ii) Adiabatic deceleration or acceleration depending on the sign of the divergence of the expanding solar wind.;
and iii) Diffusion caused by the irregular heliospheric magnetic field. The diffusion coefficient depends on one's position, on the particles' rigidity (or energy), and on the solar activity (time).;
iv) Gradient and curvature drifts in the global heliospheric magnetic field, where the drift effects depend not only on the particles' rigidity and charge but also on the phase of the 22-year solar-activity cycle, \textit{i.e.} the orientation of the solar magnetic field above the solar poles \citep[$A>0$ or $A<0$:][]{web88}.
\newline
At 1~AU the different modulation processes manifest themselves in the shape of the measured energy spectra, \textit{i.e.} in the range above 100~MeV~nucleon$^{-1}$ to below a few GeV~nucleon$^{-1}$ and its temporal variations \citep{potgieter2013}. Thus, in order to investigate these effects in more detail, cosmic-ray energy spectra covering the above-mentioned energy range at all different phases of the 22-year solar magnetic cycle are required. Furthermore, the energy coverage of measurements inside the Earth's magnetosphere depends on the position of the observer, due to the geomagnetic cutoff. Thus an instrument in interplanetary space is preferred. Since currently there is no dedicated instrumentation available, we extended the measurement capabilities of the \textit{Electron Proton Helium INstrument} (EPHIN) onboard the \textit{SOlar and Heliospheric Observatory} (SOHO), which was launched in 1995 and has been located at the Lagrangian point L$_{1}$ since 1996.\newline
After a brief synopsis of the method and an error analysis, the method is validated in comparison to various other missions. Finally, annual proton spectra from 1995 to 2014 are presented.

\section{Instrumentation}
\begin{figure}
\includegraphics[width=0.6\textwidth]{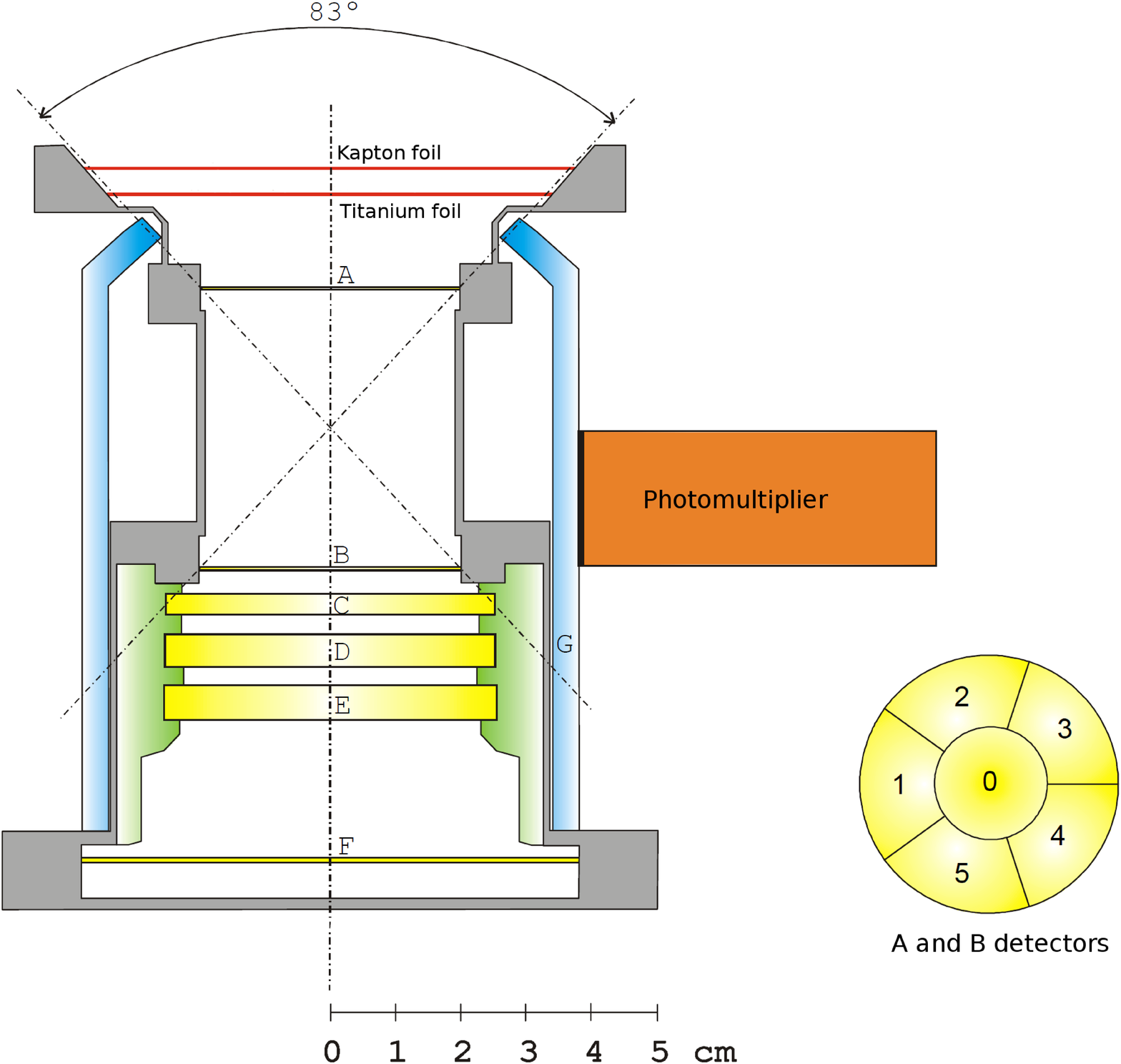} 
\caption{Sketch of the EPHIN instrument \cite[adapted from][]{gom03}.}
\label{fig:ephin}
\end{figure}
A sketch of the EPHIN instrument \citep{mue95} is shown in Figure \ref{fig:ephin}. The instrument consists of six silicon solid-state detectors surrounded by a scintillator as anti-coincidence. For particles that deposit their entire energy in the detector stack (``stopping particles"), the type of the particle measured can be easily identified using the $dE\mathrm{/}dx-E$ method \citep{mue95}. However, this method is limited for protons to energies below 50 MeV. Above that energy, protons penetrate through the instrument, depositing only a fraction of their total kinetic energy. In order to overcome this limitation, a new method previously only used during solar events \citep{kue15a} is adapted in such way that GCR proton spectra in the energy range from 250~MeV up to 1.6~GeV can be obtained.

\section{Method}
\begin{figure}
\includegraphics[width=0.8\textwidth]{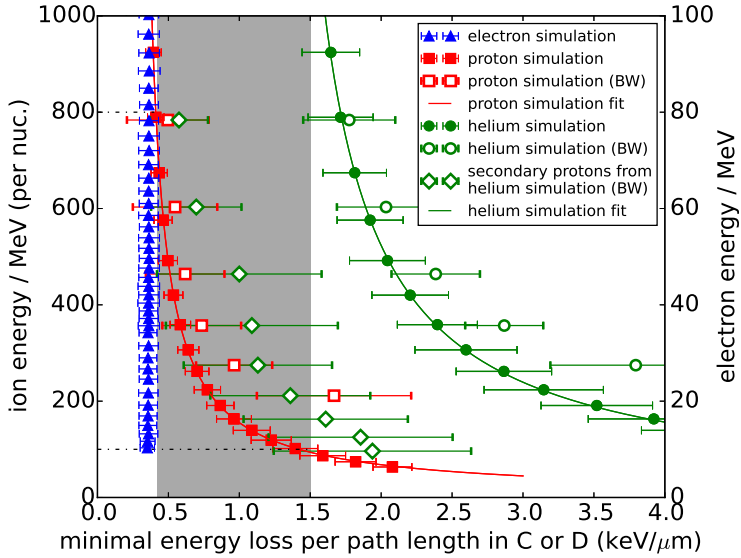} 
\caption{Relation between initial total energy and energy loss in the detector \cite[adapted from][]{kue15a}.}
\label{fig:method}
\end{figure}
The method applied \citep[for more details see][]{kue15b,kue15a} relies on energy losses of particles that penetrate the entire instrument (\textit{e.g.} detector A \,--\, F are triggered, \textit{\textit{c.f.}} Figure \ref{fig:ephin}). Sophisticated Geant4 Monte-Carlo simulations \citep{geant} have been performed for electrons, protons and helium ions. In addition to particles entering the instrument at detector A and exiting at detector F (``forward direction"), particles coming from behind the instrument (entering at F, exiting at A; ``backward direction") are also considered. For the latter, the shielding of the SOHO spacecraft is taken into account by placing a 10~cm Al layer behind the instrument. The simulation results are summarized in Figure \ref{fig:method}, where the simulated energy is presented as a function of the minimum energy loss in either the C or D detector. While the energy loss of electrons (blue triangles) is below $\approx$~0.4~keV~$\mu$m$^{-1}$ independent of their energy, helium particles (green circles) typically lose more than $\approx$~1.5~keV~$\mu$m$^{-1}$ and hence intermediate-energy losses are almost entirely caused by protons (red squares). Furthermore, the relation between total kinetic energy and energy loss for forward protons can be described by an analytical function fitted to the simulation results. Using this function, the total kinetic energy of a proton with a given measured energy loss can be estimated. However, especially at lower energies, protons passing the instrument in the backward direction differ significantly from forward penetrating particles. In addition, secondary protons can be created by backward directed helium in the shielding (green diamonds).\newline
\begin{figure}
\includegraphics[width=0.8\textwidth]{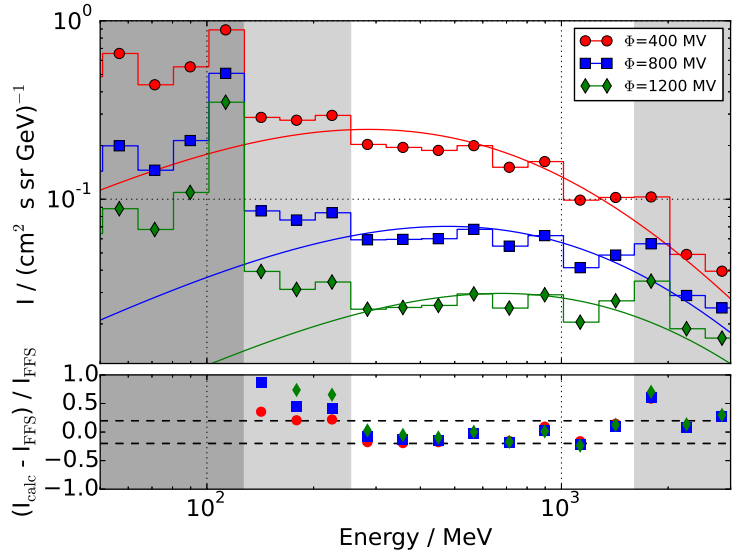} 
\caption{Input proton spectra (FFS, solid lines) and resulting spectra for different solar modulation conditions. In the bottom panel, the relative deviation between input and model spectra is shown.}
\label{fig:ffs}
\end{figure}
In order to determine the energy range in which the method can be applied and to estimate the systematic errors, further simulations using realistic proton and helium spectra as simulation input have been performed. The force field solution \citep[FFS:][]{gle68,uso11} approximates the GCR spectra at Earth for a given local interstellar spectrum (LIS) and particle type as a function of a single variable, the modulation parameter [$\phi$]. The solid lines in Figure \ref{fig:ffs} show FFS proton spectra for different $\phi$, representing different solar modulation conditions (\textit{e.g.} $\phi$=400~MV for solar minimum, $\phi$=1200~MV for solar maximum). Furthermore, Helium spectra with the same $\phi$ and a energy independent He/p ratio of 25\,\% were included in the simulation. The artificial data were then analyzed with the method described. The resulting spectra are shown as symbols in Figure \ref{fig:ffs}. In the lower panel, the relative deviation between the resulting and the input spectra are presented. Below 130~MeV (marked by dark-grey shading in the figure), high-energy helium ions cause energy loss similar to that of low-energy protons and hence the intensity is heavily overestimated \citep[Figure \ref{fig:method}, \textit{cf.}][]{kue15b}. In the energy range between 130 and 250~MeV (light grey), the intensity is slightly overestimated due to the influence of backward-directed protons \citep{kue15a}. Above 1.6~GeV (light grey), the energy loss of different energies converges and thus, without any further corrections, the errors increase. However, based on the bottom panel of Figure \ref{fig:ffs}, the systematic errors between 250~MeV and 1.6~GeV do not increase above 20\,\% for all solar modulation conditions.\newline
The statistical errors are of the order of $\approx$ 10\,\%, 2\,\% and 0.5\,\% for a spectrum of a given day, month and year respectively.\newline
\begin{figure}
\includegraphics[width=0.8\textwidth]{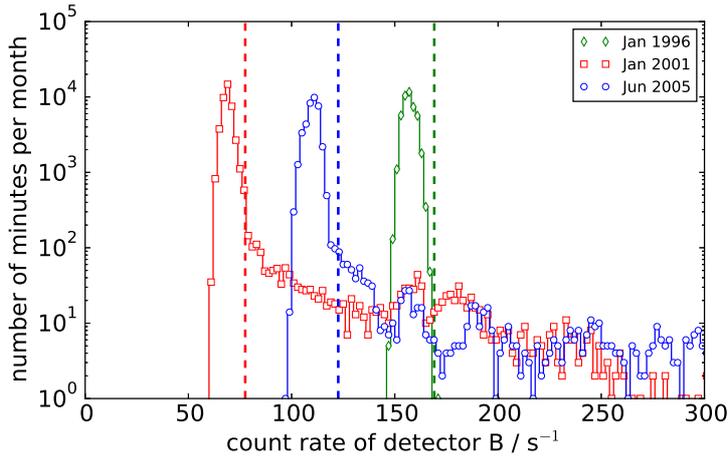} 
\caption{Example histograms of count rates of detector B for three selected months. The dashed lines show the threshold for solar events.}
\label{fig:solarevent}
\end{figure}
In order to derive the GCR spectrum in a given time interval, solar events have to be excluded from the data set as they feature higher fluxes and a different spectral shape \citep{mewaldt2012}. Therefore, we analyzed the count rate of detector B (\textit{cf.} Figure \ref{fig:ephin}) without any coincidence condition, which is available at one-minute resolution. The B-detector is sensitive to particles of low energy \citep{kue15c} and hence the count rate is expected to rise during solar events. In Figure \ref{fig:solarevent}, histograms of the detector B count rate are shown for three different months. All three histograms feature a sharp peak at count rates between 50 and 170 counts per second. Those count rates are caused by the GCR and the variation of the position of the peak can be explained by solar modulation \citep{kue15c}. In addition, the histograms for January 2001 and June 2005 show the occurrence of count rates above 170 counts per second due to solar events. In our analysis, we have fitted a Gaussian to the peak of the count-rate distribution for every single month and excluded time periods, in which the count rates in detector B rose above 3$\sigma$ over the mean of the Gaussian (dashed lines in Figure \ref{fig:solarevent}). \newline
Following a similar approach to that used by \citet{kue15a}, \citet{morgado2015} have analyzed the response of the \textit{Electron, Proton and Alpha Monitor} (EPAM) onboard the \textit{Advanced Composition Explorer} (ACE) and the \textit{Heliosphere Instrument for Spectra, Composition and Anisotropy at Low Energies} (HISCALE) onboard \textit{Ulysses} to penetrating particles, focussing on the possibility of detecting anisotropies and onset times of the May, 2012 GLE for particles above 1~GeV. However, since those instrument do not have an integral coincidence channel and thus a clearly defined direction of arrival as well as data of the energy loss in several detectors, the analysis of penetrating particles is much more complicated and an energy spectrum for penetrating particles can not be derived.

\section{Results and Conclusion}
\begin{figure}
\includegraphics[width=0.8\textwidth]{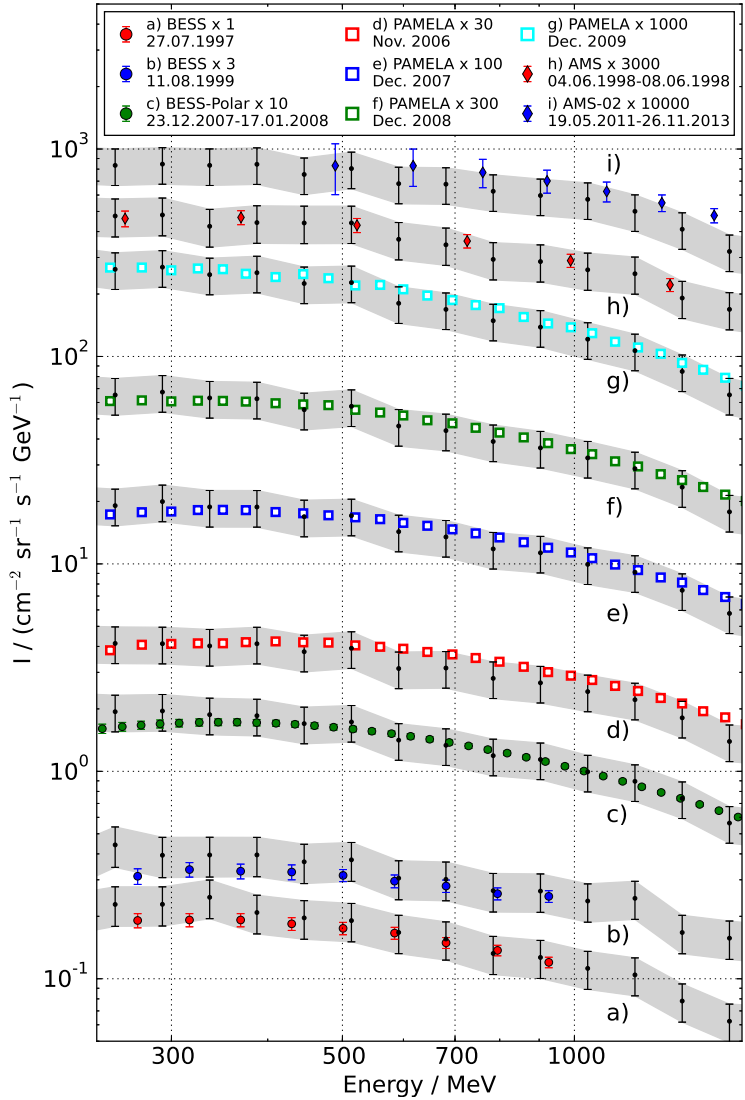} 
\caption{Proton spectra from AMS, AMS-02, BESS, BESS-Polar, and PAMELA in comparison to the derived EPHIN spectra (black). Spectra are scaled in the interest of greater clarity.}
\label{fig:spectra}
\end{figure}
In order to validate the derived spectra, Figure \ref{fig:spectra} shows data from various balloon missions (\textit{Balloon-borne Experiment with Superconducting Spectrometer} (BESS): \citealp{bess}, and BESS-Polar: \citealp{bess-polar}) as well as space-borne instruments (AMS: \citealp{ams01}, AMS-02: \citealp{ams02}, and PAMELA: \citealp{pamela}) in comparison to EPHIN spectra (black) for the same time periods. The different spectra are scaled in the interest of greater clarity. The square sums of the statistical and the systematic error (20\,\% independent of energy) of EPHIN are shown as an error band (grey). Note that some BESS and BESS-Polar results are not shown, due to either gaps in EPHIN data or solar energetic-particle events in the given time period, masking the GCR spectra.\newline
Based on this figure, EPHIN spectra are in good agreement with the other measurements when taking into account the errors. Since the different time periods cover different phases of the solar cycle, it can be concluded that the method presented is valid for deriving spectra of galactic cosmic rays consistently. \newline
Based on the method presented, annual GCR proton spectra from 1995 until 2014 have been derived. All annual spectra are available in the Appendix. \newline
\begin{figure}
\includegraphics[width=0.8\textwidth]{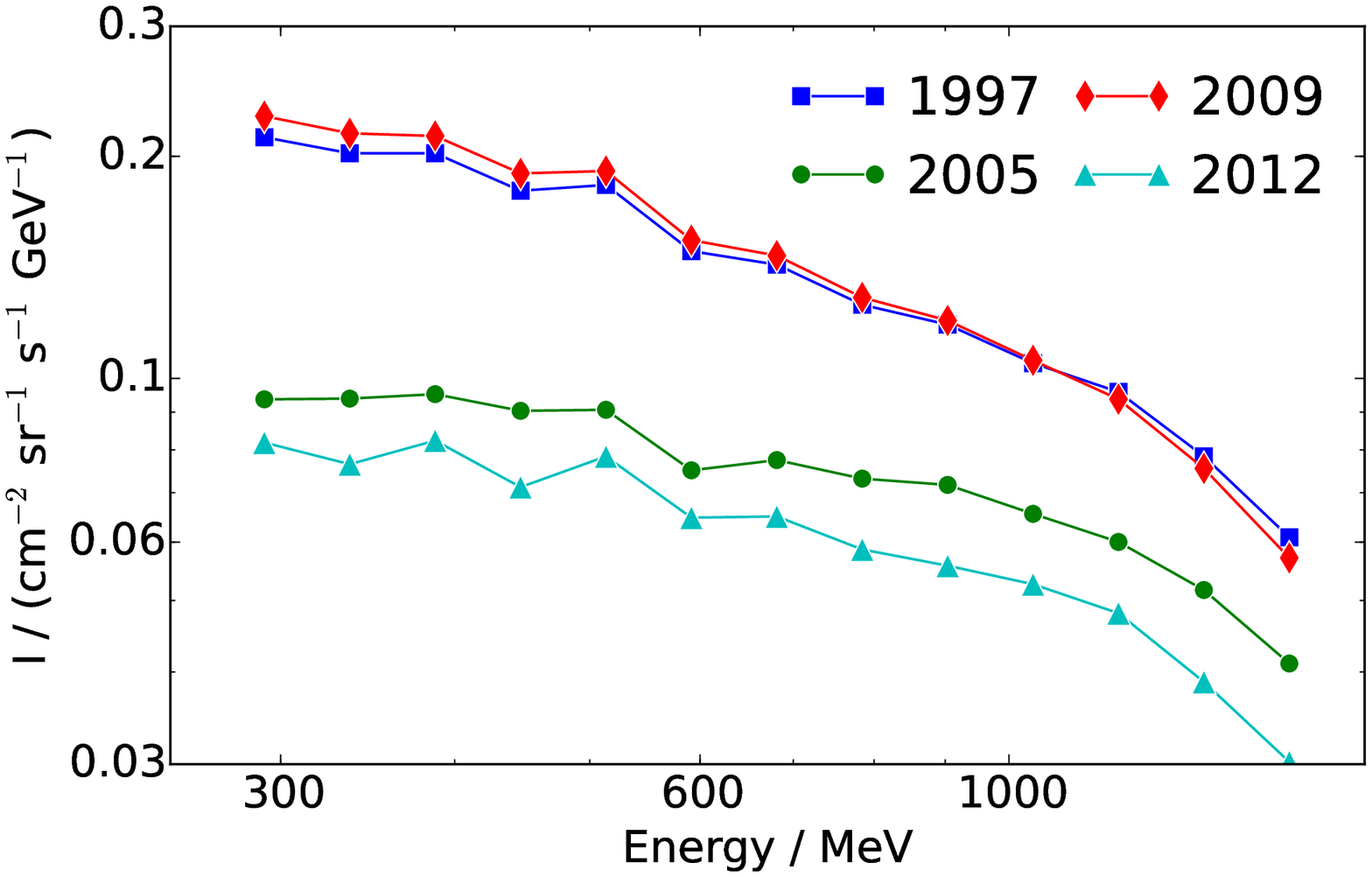} 
\caption{A selection of annual proton spectra derived from SOHO/EPHIN data with the method presented.}
\label{fig:annual_spectra}
\end{figure}
\begin{figure}
\includegraphics[width=0.8\textwidth]{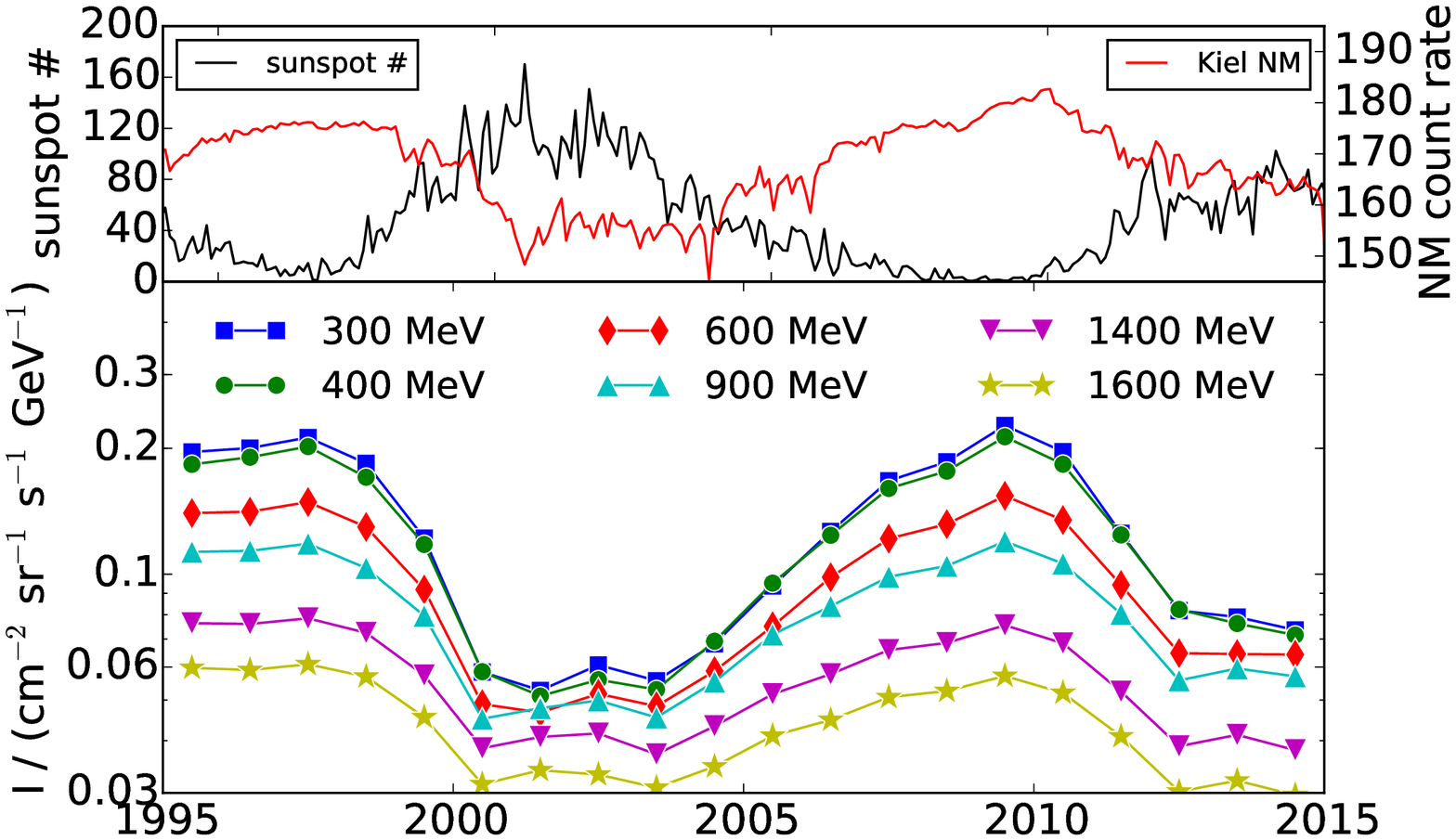} 
\caption{Upper panel: Monthly averaged sunspot number (black, left-hand axis), Kiel neutron-monitor count rate (red, right-hand axis). Lower panel: Proton intensity variations at different energies over the last two decades. The data were derived from SOHO/EPHIN data with the method presented.}
\label{fig:annual_channels}
\end{figure}
A selection of spectra at different phases of the solar cycle is presented in Figure \ref{fig:annual_spectra}. While the two spectra at solar minimum (1997 and 2009) were obtained during different polarities ($A>0$ and $A<0$ respectively) the other two spectra (2005 and 2012) were obtained during the declining and rising phase of a $A<0$ polarity phase \citep{hat10}. Especially the ``record-high" spectrum of 2009 \citep{mew10} represents a particular challenge for models to explain. In order to investigate the solar cycle dependence in more detail, Figure \ref{fig:annual_channels} displays the monthly averaged sunspot number (\url{solarscience.msfc.nasa.gov/SunspotCycle.shtml}) and the count rate of the Kiel neutron monitor (\url{www.nmdb.eu}, both in the upper panel) in comparison to the yearly averaged derived intensities at selected energies between 300 and 1600~MeV (lower panel). Both the neutron-monitor count rate and the intensities at different energies measured by SOHO/EPHIN are anti-correlated with the solar activity represented by the sunspot number \citep{heber2006a, heber2006b}. Furthermore, the intensity variation due to the solar modulation increases with decreasing energy. This behaviour is expected based on Parker's transport equation \citep{potgieter2013}. Note that similar to the neutron-monitor count rate, the intensity variations show features of drift effects such as the sharp peak in 2009 ($A<0$) in contrast to the flatter maximum in 1997 ($A>0$) \citep{web88}.\newline
Both figures show that the new data set allows us to investigate modulation processes at energies below the one obtained by neutron monitors and above the usual energy range from spacecraft instrumentation. Thus the two decades of data available, together with the unique position of SOHO outside the Earth's magnetosphere, offer the opportunity to validate solar-modulation model studies \citep[\textit{e.g.}][]{pot14}.

\acknowledgments
The SOHO/EPHIN project is supported under Grant 50~OC~1302 by the German Bundesministerium f\"ur Wirtschaft through the Deutsches Zentrum f\"ur Luft- und Raumfahrt (DLR).\newline
This work was carried out within the framework of the bilateral BMBF-NRF-project ``Astrohel" (01DG15009) funded by the Bundesministerium f\"ur Bildung und Forschung. The responsibility of the contents of this work is with the authors.\newline
This project has received funding from the European Union's Horizon 2020 research and innovation programme under grant agreement No 637324.\newline
Ra\'ul G\'omez-Herrero acknowledges the financial support by the Spanish \linebreak MINECO under projects ESP2013-48346-C2-1-R and AYA2012-39810-C02-01.

\section*{Disclosure of Potential Conflicts of Interest}
The authors declare that they have no conflicts of interest.

\begin{table}
\begin{tabular}{ c | c  c  c  c  c  c  c  c  c  c }
\hline
E \textbackslash Y & 1995 & 1996 & 1997 & 1998 & 1999 & 2000 & 2001 & 2002 & 2003 & 2004 \\ 
\hline
292 & 0.196 & 0.200 & 0.212 & 0.184 & 0.122 & 0.058 & 0.053 & 0.061 & 0.056 & 0.068 \\ 
336 & 0.183 & 0.190 & 0.202 & 0.171 & 0.115 & 0.056 & 0.051 & 0.059 & 0.052 & 0.068 \\ 
387 & 0.183 & 0.190 & 0.202 & 0.171 & 0.118 & 0.058 & 0.051 & 0.056 & 0.053 & 0.069 \\ 
446 & 0.169 & 0.171 & 0.180 & 0.154 & 0.107 & 0.051 & 0.049 & 0.054 & 0.050 & 0.064 \\ 
513 & 0.170 & 0.174 & 0.183 & 0.158 & 0.111 & 0.055 & 0.054 & 0.059 & 0.055 & 0.067 \\ 
591 & 0.140 & 0.141 & 0.149 & 0.130 & 0.092 & 0.049 & 0.047 & 0.052 & 0.048 & 0.059 \\ 
681 & 0.135 & 0.136 & 0.143 & 0.125 & 0.090 & 0.050 & 0.046 & 0.055 & 0.049 & 0.060 \\ 
784 & 0.119 & 0.121 & 0.126 & 0.111 & 0.082 & 0.046 & 0.046 & 0.051 & 0.047 & 0.056 \\ 
903 & 0.113 & 0.114 & 0.118 & 0.103 & 0.080 & 0.045 & 0.048 & 0.050 & 0.045 & 0.055 \\ 
1040 & 0.101 & 0.101 & 0.105 & 0.096 & 0.073 & 0.046 & 0.048 & 0.049 & 0.046 & 0.054 \\ 
1198 & 0.094 & 0.093 & 0.096 & 0.087 & 0.068 & 0.043 & 0.045 & 0.046 & 0.043 & 0.050 \\ 
1380 & 0.076 & 0.076 & 0.078 & 0.072 & 0.057 & 0.038 & 0.041 & 0.042 & 0.037 & 0.043 \\ 
1589 & 0.060 & 0.059 & 0.061 & 0.057 & 0.046 & 0.031 & 0.034 & 0.033 & 0.031 & 0.035 \\ 
\hline
\hline
E \textbackslash Y & 2005 & 2006 & 2007 & 2008 & 2009 & 2010 & 2011 & 2012 & 2013 & 2014 \\ 
\hline
292 & 0.094 & 0.127 & 0.167 & 0.186 & 0.227 & 0.196 & 0.125 & 0.082 & 0.079 & 0.074 \\ 
336 & 0.094 & 0.122 & 0.159 & 0.176 & 0.215 & 0.185 & 0.126 & 0.077 & 0.076 & 0.072 \\ 
387 & 0.095 & 0.124 & 0.160 & 0.176 & 0.213 & 0.183 & 0.124 & 0.082 & 0.076 & 0.072 \\ 
446 & 0.090 & 0.113 & 0.144 & 0.159 & 0.190 & 0.163 & 0.109 & 0.071 & 0.071 & 0.067 \\ 
513 & 0.091 & 0.117 & 0.147 & 0.160 & 0.191 & 0.166 & 0.116 & 0.078 & 0.075 & 0.073 \\ 
591 & 0.075 & 0.098 & 0.122 & 0.132 & 0.154 & 0.135 & 0.094 & 0.065 & 0.064 & 0.064 \\ 
681 & 0.078 & 0.097 & 0.117 & 0.126 & 0.147 & 0.129 & 0.092 & 0.065 & 0.065 & 0.064 \\ 
784 & 0.073 & 0.086 & 0.104 & 0.112 & 0.129 & 0.114 & 0.083 & 0.059 & 0.062 & 0.061 \\ 
903 & 0.072 & 0.084 & 0.099 & 0.105 & 0.120 & 0.106 & 0.080 & 0.056 & 0.059 & 0.057 \\ 
1040 & 0.066 & 0.076 & 0.088 & 0.093 & 0.106 & 0.097 & 0.075 & 0.053 & 0.058 & 0.058 \\ 
1198 & 0.060 & 0.070 & 0.080 & 0.085 & 0.094 & 0.085 & 0.065 & 0.048 & 0.049 & 0.046 \\ 
1380 & 0.052 & 0.058 & 0.066 & 0.068 & 0.076 & 0.068 & 0.052 & 0.039 & 0.041 & 0.038 \\ 
1589 & 0.041 & 0.045 & 0.051 & 0.053 & 0.057 & 0.052 & 0.041 & 0.030 & 0.032 & 0.030 \\ 
\hline
\end{tabular}
\caption{Annual proton spectra from 300~MeV up to 1.6~GeV from 1995 to 2014. Energy is given in MeV, differential intensity in (cm$^2$\ sr\ s\ GeV)$^{-1}$. The systematic errors are approximated to be less than 20\,\%, statistical errors are less than 1\,\%.}
\label{tab:spec}
\end{table}

\clearpage

\bibliographystyle{spr-mp-sola}
\bibliography{kuehl_et_al}

\end{article} 
\end{document}